\documentclass[aps,preprint,superscriptaddress,floatfix,showkeys]{revtex4-1}

\usepackage[pdftex]{graphicx}
\usepackage{dcolumn}
\usepackage{bm}

\usepackage{amssymb}
\usepackage{multirow}

\begin{document}

\title{Heuristic Monte Carlo Method Applied to Cooperative Motion Algorithm for Binary Lattice Fluid}

\author{P. Knycha{\l}a}
\email[]{p.knychala@pwsz.kalisz.pl}
\affiliation{Chair of Computer Science, The President Stanislaw Wojciechowski Higher Vocational State School in Kalisz \\
ul. Nowy Swiat 4, 
62-800 Kalisz, 
Poland}

\author{M. Banaszak}
\affiliation{ Faculty of Physics,
A. Mickiewicz University \\
ul. Umultowska 85,
61-614 Poznan,
Poland}

\date{\today}

\begin{abstract}
The Cooperative Motion Algorithm is an efficient lattice method to simulate dense polymer systems and is often used with two different criteria to generate a Markov chain in the configuration space. 
While the  first method is the well-established  Metropolis algorithm, the other one is an heuristic algorithm which needs
justification. As an introductory  step towards justification for the  3D lattice polymers, we study a simple system which is the binary equimolar fluid on a 2D triangular lattice. Since all lattice sites are occupied only selected type of motions are considered, such the vacancy movements, swapping neighboring lattice sites (Kawasaki dynamics)  and cooperative loops. 
We compare both methods, calculating the energy as well as heat capacity as a function of temperature. 
The critical temperature, which was determined using the Binder cumulant, was the same for all methods with the simulation accuracy and in agreement with the exact critical temperature for the Ising model on the $2D$ triangular lattice. In order to achieve reliable results at low temperatures we employ the parallel tempering algorithm 
which enables  simultaneous simulations of replicas of the system in a wide range of temperatures. 

\end{abstract}


\maketitle

\section{INTRODUCTION}
Computer simulations are one of the most important research tools in modern science~\cite{Frenkel2001}. In particular, 
Monte Carlo method, which  generates a Markov chain in the space of states, is used in physics and chemistry problems since 1953~\cite{Metropolis1953}. 
In this algorithm, various states of system are generated in such a way and the next state depends exclusively on the previous state. A random walk in the state space can be generated in order to achieve a given density of states in such a manner that some new states are accepted and  the other ones rejected. 

One of the most celebrated  examples (of using a lattice Monte Carlo method) is the Ising model which was developed for magnetic systems~\cite{Binder1997, Newman1999}. In this model each spin can be directed up or down and the number of the up and down spins is not preserved, so the order parameter (total magnetization) is not conserved. 
The Ising model with conserved order parameter can also be employed to study the separations in liquid mixtures~\cite{Fratzl1994, Halagan2009}. 

Monte Carlo method is also used to investigate properties of polymers. Those materials are increasingly used in many areas of life and have the fascinating ability to self assemble into periodic nanostructures. Polymer systems consist of many macromolecules and each one is composed of numerous repeat units, that are called monomers or segments. Computer simulations of such large system are not trivial so that different optimizations and simplifications are used.
For example, in coarse-grained methods details of the chemical structure of polymer are neglected and a few chemical monomers are replaced by single segment. Another simplification used in computer simulations of polymer is the use of lattice, where most floating-point calculations are eliminated. 

A simulation of dense lattice polymer models is a considerable challenge because the lattice sites are completely filled with chain segments and two segments cannot occupy the same space, and therefore it is not a straightforward task to generate a Markov chain in the configuration space. 
For lattice fluid,
the Kawasaki dynamics~\cite{Kawasaki1989} can be used,  in which random pairs of neighboring spins are swapped. 
This method, however, cannot be easily extended to polymer systems, because the linear connectivity is usually
violated by random swapping. 
An alternative approach is the vacancy dynamics in which the Monte Carlo movements  are driven by diffusion of a single vacancy~\cite{Fratzl1994}. There are also other approaches, using more than one vacancy, proposed by Yaldram and Binder~\cite{Yaldram1991, Yaldram1991b}. Finally, there are methods which use cooperative loops, either in small 
incremental steps in a single cooperative loop. For lattice polymer,
 Pakula developed  two methods (employing cooperative loops, one way or the other)  to generate the Monte Carlo  movements: Cooperative Motion Algorithm~\cite{Pakula1987A, Pakula1987B, Pakula1988} (CMA) and Dynamic Lattice liquid (DLL) Algorithm~\cite{Pakula1997}.

The CMA seems to be the most efficient method to simulate dense polymer system. Representative states of system are generated by cooperative movements of segments along the random walk loops on the face-centered cubic lattice (FCC). 
The FCC lattice is very suitable for simulation of dense polymer system since the coordination number, $z$, of this lattice is $12$, which is similar to liquid where average number of nearest neighbors is about $12$.
Acceptance of the new state of the system can be achieved by using well  known Metropolis rules~\cite{Metropolis1953} or using a heuristic method proposed by Pakula~\cite{Pakula1993}. In the first case, the difference between energy of new and previous states is considered, and in case of heuristic method only the energy of new state  is taken into account.
The CMA method was used to investigated various polymer system
~\cite{Weyersberg1993, Gauger1993, Pakula1996, Pakula1997B, Banaszak2002, Qin2003, Majtyka2006, Jeszka2006, Polanowski2007, Majtyka2007, Sikorski2011, Polanowski2013, Knychala2014} but the comparison of the different criteria of state acceptance (the Metropolis and heuristic approach) were not systematically  studied and described in literature. 
For example, Vilgis and Weyersberg~\cite{Weyersberg1993} applied CMA with the Metropolis test to analyze phase behavior of symmetric diblock copolymer and their results were favorably compared with random-phase approximation.
The same version of CMA was also used to investigate static properties of a dendrimer with positively charged terminal groups by Majtyka and K\l{}os~\cite{Majtyka2007}.
The CMA with heuristic method was employed to investigate properties of copolymer system with various distributions of monomers along the chain~\cite{Pakula1996}. Static and dynamic properties of diblock copolymer melts near the order - disorder transition were investigated using heuristic method in ref.~\cite{Pakula1997B}.
Both above methods were used to determine microphase separation in triblock copolymers~\cite{Banaszak2002} and  in ion-containing block copolymers~\cite{Knychala2014}.

In this paper, we compare two methods: 
\begin{itemize}
	\item standard Metropolis (with Kawasaki dynamics and  cooperative loops)
	\item heuristic method (with cooperative loops)
\end{itemize}
The heuristic method was often used with CMA, but to best of our knowledge, has not been systematically compared
with the Metropolis algorithm.
We use simple $2D$ triangular lattice model of liquid consisting of two type of elements which is similar to the Ising model (however the order parameter is conserved) instead of complex polymer system and seek to answer the following questions:
\begin{enumerate}
\item How different criteria of new state acceptance and Monte Carlo movements (full loop vs. loops divided into
incremental steps)  influence the critical temperature below which the system separates into two phases?
\item  How the above criteria  influence the energy or heat capacity; are they quantitatively consistent?
\end{enumerate}

\section{SIMULATION METHODS}

Since the main aim of this paper is to investigate algorithms, we used simple $2D$ triangular lattice liquid instead of $3D$ lattice. 
The lattice of size $L \times L$ consists of $L^2/2$ sites and each lattice site contains one of two types of segments: $A$ or $B$. The volume fraction $f$ of segment $A$ is equal to $0.5$ and for segment $B$ it is $1-f$. The number of each type of segments is constant during simulations so that it is a model with conserved order parameter. 
An attempt to move a single segment defines a single Monte Carlo step.
There are no vacancies on the lattice so that segments can move via cooperative motion in which a few elements are moved at the same time, as shown in fig.~\ref{figcma}. 
This single cooperative motion is called a cooperative loop. Figure~\ref{figcma}a presents a single loop before segments are moved and the final state is presented in fig.~\ref{figcma}b. 
In the athermal state (without interaction between segments) this cooperative loop is created in the following way:
\begin{enumerate}
	\item a segment from random lattice site is removed (temporary vacancy is created) and this site is labeled both INITIAL and CURRENT,
	\item one of six neighbors of CURRENT lattice site is selected at random,
	\item segment from this selected site is moved to CURRENT site and selected site becomes the CURRENT one,
	\item points 2 and 3 are iterated until the random neighbor is not the same as the INITIAL site,
	\item the segment which was initially removed is put into the vacancy.
\end{enumerate}
After athermal simulation, consisting of such cooperative loops, the segments are randomly distributed on the lattice.

Next, interactions between segments are introduced and described through $\epsilon_{ij}$ parameters. The interaction between different type of segment $\epsilon_{AB}$ is equal to $\epsilon$ and between the same type of segment is $0$ ($\epsilon_{AA}=\epsilon_{BB}=0$). The interaction are limited to nearest neighbors and energy of single segment $i$ is $E^*_i = \sum_{j=1}^{6}\epsilon_{ij}$ ($\epsilon$ parameter is also used as energy unit). The reduced temperature $T^*$ is $kT/\epsilon$. It means that segments prefer segments of the same type and below the critical temperature, $T_c$, the system separates in two phases, first is rich in segments $A$ and the second in segments $B$.

In case of this thermal simulation various states of system are generated using the same cooperative motions (and also Kawasaki dynamics motions) and probability transition between current state $X$ and new state $X'$ is described by following formula:
\begin{equation}
	P(X\rightarrow X') = min\left [ 1, \exp\left ( -\frac{\Delta E}{kT} \right ) \right ].
\end{equation}
In this paper, three approaches employing the Metropolis criterion and also the heuristic method are used. 

In the first case, the loop is created using single vacancy which moves as a random walk. The difference of energy is calculated for full loop and $\Delta E = E_{X'} - E_X$, where $E_X$ is energy of the initial state (state before cooperative loop) and $E_{X'}$ is energy of the state after cooperative loop. We use standard Metropolis criteria~\cite{Metropolis1953}, and refer to this method as $GM$, the motions are non-local since more then one segment is moved in a single step of simulation. In this case some mismatch with theory can occur due to possible large energy changes involved with acceptance of long loops, as discussed in next chapter. 

In the second case ($LM$), the loop is divided into incremental steps in which the movement of a single segment is analyzed. We refer to it as Metropolis with single vacancy. At the beginning of the loop, the segment from random lattice site is removed without any condition since it decrease energy of system (vacancy which does not interact with segments is created). At the end of the loop, the removed segment is inserted with probability proportional to Boltzmann factor since this motion always increases energy of system. In case of failure, the whole loop is rejected. Inside the loop, $E_X$ describes energy of sliding segment at initial position as shown in fig~\ref{metropolis}a and $E_{X'}$ describes energy of the same segment at new position (fig.~\ref{metropolis}b). In a single step, we try to move only one segment.

Moreover, the Metropolis method was employed for the Kawasaki dynamics~\cite{Kawasaki1972} ($KA$), swapping neighboring lattice sites, which generated results that we use as a background for other methods.

The last approach is similar to $LM$ method with one exception, the acceptance condition is always based on energy of single segment in new position (energy of previous state is not included in Boltzmann factor). It means that $\Delta E = E_{X'}$.
We call this method as heuristic Monte Carlo method ($HE$) since the correctness of this approach is not obvious and compare it to commonly used Metropolis criterion can be interesting and provide additional and intriguing insights into CMA algorithm.

We use five different lattice size in presented simulations: $40\times 40$, $50\times 50$, $60\times 60$, $80\times 80$, and $100\times 100$. 
First of all, the athermal simulations is performed for at least $2\times 10^6$ MCS. Next, we perform thermal run from $3\times 10^6$ to $6\times 10^6$ MCS depending on lattice size and the first half of run is needed to equilibrate the system, and the second half is used to collect the data. All simulation experiments are repeated at least 3 times starting with different initial athermal states.

In order to improve statistics (particularly at low temperatures) the parallel tempering~\cite{Earl2005, Swendsen1986} (PT) algorithm was also used in simulations. In the PT method, $M$ replicas of system are simulated in parallel (in this case it was $48$), each in different temperature $T^*_i$, with $i$ ranging from $1$ to $M$. After a number of Monte Carlo steps (MCS) (in our case it was 500 MCS),
we try to exchange replicas with neighboring $T^*_i$ in random order with the following probability:
\begin{equation}
	P(R_i\rightarrow R_{i+1}) = min\left [ 1, \exp\left ( -( \beta_i-\beta _{i+1} )(E_{i+1}-E_i) \right ) \right ] 
\end{equation}
where $\beta_i = 1/kT$. 

\section{RESULTS AND DISCUSSION}

First, we carry out simulations for lattice size equal to $50\times 50$. In fig.~\ref{fig1}, energy per lattice site, $E^*$, and heat capacity, $C_v$ in terms of temperature for different methods are presented.
The heat capacity is calculated from the equation:
\begin{equation}
	C_v = \frac {\left \langle \left ( E^* - \left \langle E^* \right \rangle \right )^2 \right \rangle} {kT^{*2}}
\end{equation}
One can observe single peak in $C_v$ at $1/T^* = 0.55$ ($T^* = 1.82$) for all Metropolis approaches as well as heuristic method which is related to the separation of the system into two phases. Above this critical temperatures, $T_c$, the segments of type $A$ and $B$ are uniformly distributed on the lattice (like in fig~\ref{fig3}b), while below $T_c$ system start to separate into two phases (fig~\ref{fig3}a). The first one is rich in segments of type $A$ and the second one mostly consists of segments $B$. A significant decrease in energy is observed also at $T_c$, as shown in fig.~\ref{fig1}b. This temperature $T_c$ is consistent with the theoretical value of critical temperature for the Ising model at $2D$ triangular lattice (with coordination number $z=6$) calculated by Houtappel in 1950~\cite{Houtappel1950} and equals to $1/\ln(\sqrt{3}) = 1.821$.

In fig.~\ref{fig2}a, we present probability of state acceptance, $P(X)$, in term of loop length, $l$, for the $GM$ method. It is obvious that this probability decreases with loop length increasing and for loop length greater than $200$ is close to $0$. We introduces additional parameter that describes contribution of different loop lengths. 
It determines what length of the loop takes most time of simulation and is described as:
\begin{equation}
	 C(l) = l P(l),
\end{equation}
where $P(l)$ is the probability of the loop of length $l$. This parameter is shown in fig.~\ref{fig2}b. 
The contribution decreases with loop length increasing and reaches a minimum at about $200$, and then begins to rise sharply. The contribution of short loop length (shorter than $50$) is significant but the contribution of long loops (greater than $200$) is more significant. As shown in fig.~\ref{fig2}a the probability of acceptance of long loops is close to $0$ and this affects the efficiency of $GM$ method. Since the $GM$ method is the least efficient, requiring the longest time of simulation, we do not take this  approach into account in further analysis.

Moreover, it is possible to determine of the critical temperature with greater precision by calculating the Binder cumulant~\cite{Binder1992} ($U_L$):
\begin{equation}
\label{order_parameter}
U_L = 1 - \frac{\left \langle (m-0.5)^4 \right \rangle}{3 \left \langle (m-0.5)^2 \right \rangle^{2}}, 
\end{equation}
where $m$ is the order parameter described by equation~\cite{Das2006}:
\begin{equation}
\label{order_parameter}
m = 2 \int_{0.5}^1 \! x_A P(x_A) \, \mathrm{d}x.
\end{equation}
The $P(x_A)$ is calculated from the histogram of the probability of finding a particle of a given type in a given area.
It can be calculated for different temperatures in terms of local concentration $x_A$ of segments type A defined as $N_A/N_R$, where $N_A$ is the number of segment type $A$ in specified area and $N_R$ is the total number of segments in the same area. In contrast to the Ising model where the number of spins is not preserved, in presented simulations the volume fraction of segments of different types is constant so $P(x_A)$ can not be calculated for the whole lattice (in that case $U_L$ will be $0$). In this paper, the radius of this area depends on size of lattice and is equal to $0.3L$ as shown in fig.~\ref{fig3}. Such approach was previously used and described in ref.~\cite{Kwak2006}.
The value $0.3$ was chosen arbitrary from a few selected tested values because it provided the best statistics. Moreover, we observed that $T_c$ does not depend on this parameter.

The sample $P(x_A)$ for different temperatures are presented in fig.~\ref{fig4}a. For $T > T_c$ the $P(x_A)$ has Gaussian distribution with maximum at $x_A = 0.5$ and the peak is getting narrower as the temperature is increased. This means that, statistically, in selected area the same number of segment $A$ and $B$ is observed (fig.~\ref{fig4}d). At $T_c$ system starts to separate into two phases as shown in fig.~\ref{fig4}c.
Below critical temperature $P(x_A)$ has bimodal distribution with two maxima at $x_A = 0$ and $1$ corresponding to two  coexisting phases (fig.~\ref{fig4}b). Those maxima become narrower with decreasing temperature since the clusters grow and become more homogeneous.

The value of the Binder cumulant depends on size of lattice for different values of temperatures except of $T_c$ where $U_L$ tends towards an universal value and not depending on $L$. For $T > T_c$ segments are uniformly mixed (system is disordered) and $U_L \rightarrow 0$ while for $T < T_c$ system is separated into two phases (ordered) and $U_L \rightarrow 2/3$ as shown in fig.~\ref{fig4b}.

The simulations for different size of lattice were carried out to determine the $T_c$ using the Binder cumulant. We use four values of lattice size: $40\times 40$, $60\times 60$, $80\times 80$, and $100\times 100$ which are presented in fig.~\ref{fig4c}. 
Results of simulations for two Metropolis approaches and heuristic method are presented in fig.~\ref{fig5}-\ref{fig8}. The summary of the results is presented in table~\ref{table:results}. 
The critical temperatures presented in table~\ref{table:results} are determined on the basis of intersection of the Binder cumulant calculated for different lattice size.
We use 48 replicas, different temperatures, in parallel tempering simulations in range form $T^*=1.62$ to $2.02$. The distance between neighboring temperatures is $0.0085$ so the simulation error (accuracy of the simulation) is about $0.5\%$ ($0.0085/1.821$). In the case of Kawasaki dynamics which is used as the background for other methods, $T_c = 1.816$ and the error is $0.24\%$. For $LM$ step acceptance method the average $T_c$ is $1.826$ and the error is $0.31\%$. The $T_c = 1.823$ with error equals to $0.16\%$ is obtained for $HE$ method. The critical temperature obtained for heuristic method is consistent with both Metropolis approaches. In all cases the value of error is below the accuracy of the simulation.

\section{CONCLUDING REMARKS}

We carried out Monte Carlo simulations of simple binary liquid using two methods: 
\begin{itemize}
	\item standard Metropolis (with different approaches), 
	\item heuristic method. 
\end{itemize}
The parameters determined during simulation with different approaches are qualitatively and quantitatively consistent. The critical temperature obtained for all methods is the same with the simulation accuracy. Moreover, the Metropolis with vacancy as well as heuristic method (where loops are divided into incremental steps) are more efficient than standard Metropolis method with full loop acceptance what is important for the simulation of dense polymer system. 

\begin{acknowledgments}
P.K. and M.B. gratefully acknowledge the research grant from the Polish NCN No. DEC-2012/07/N/ST4/00293 and the 
computational grant from the Supercomputing and Networking Center (PSNC) in Poznan, Poland.
\end{acknowledgments}

\newpage

\bibliography{cma}

\newpage

\begin{table}
  \begin{tabular}{ c  c  c  c }
    \hline
    intersection & KA & LM &  HE \\ \hline\hline
    $40\times 40 - 60\times 60$ & $1.824$ & $1.815$ &  $1.809$ \\ \hline
    $40\times 40 - 80\times 80$ & $1.819$ & $1.822$ &  $1.826$\\ \hline
    $40\times 40 - 100\times 100$ & $1.816$ & $1.825$ &  $1.823$\\ \hline    
    $60\times 60 - 80\times 80$ & $1.810$ & $1.828$ &  $1.837$\\ \hline    
    $60\times 60 - 100\times 100$ & $1.814$ & $1.831$ &  $1.826$\\ \hline            
    $80\times 80 - 100\times 100$ & $1.815$ & $1.836$ &  $1.820$\\ \hline     \hline  
    average & $1.816$ & $1.826$ &  $1.823$\\ \hline            
    standard deviation & $0.005$ & $0.007$ &  $0.009$\\ \hline                
    error & $0.24\%$ & $0.31\%$ &  $0.16\%$\\ \hline                    
  \end{tabular}
  \caption{Critical temperatures for different methods obtained as intersection points of the Binder cumulant from different lattice sizes.}
  \label{table:results}   
\end{table}

\clearpage
\textbf{Figure captions:} 

\begin{figure}[ht]
\includegraphics[scale=0.5]{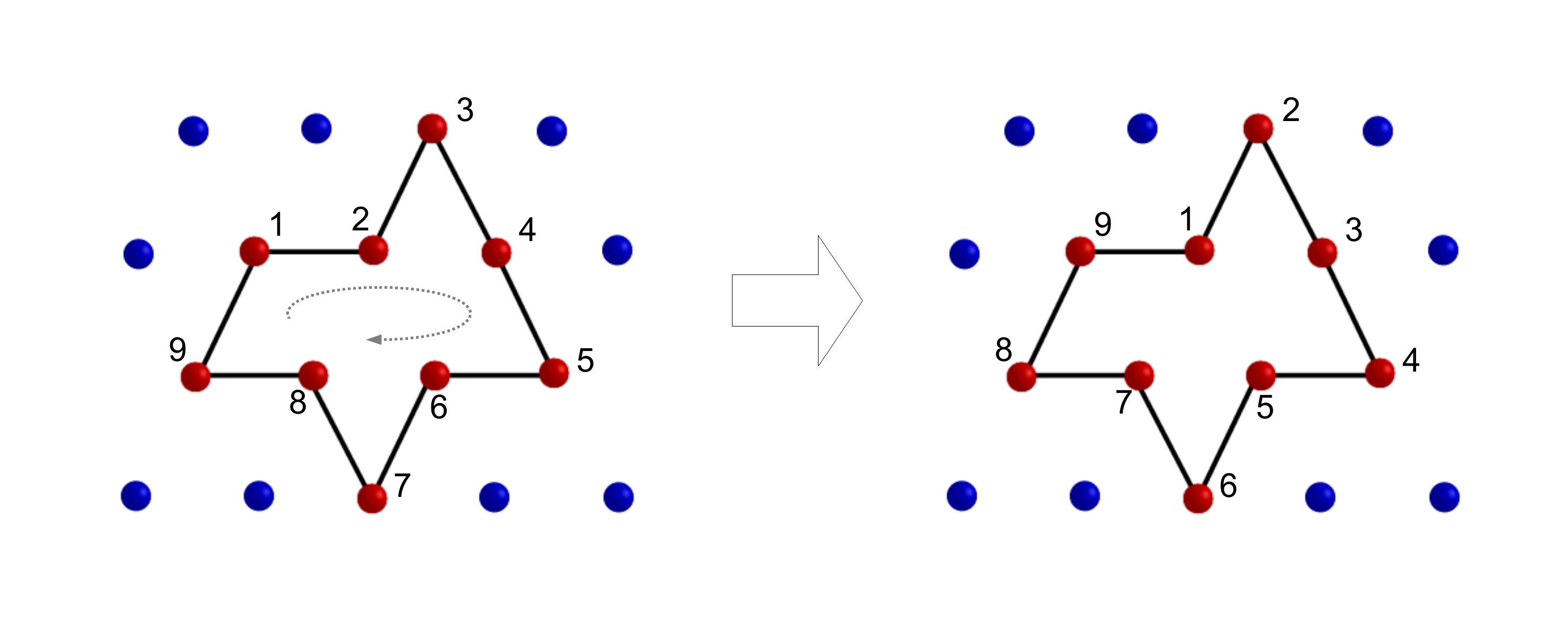} 	
\caption{ }
\label{figcma}
\end{figure}

Fig.~\ref{figcma}: Example of single cooperative motion for triangular $2D$ lattice. Grey arrow show direction of segment motion.

\begin{figure}[ht]
\includegraphics[scale=2.0]{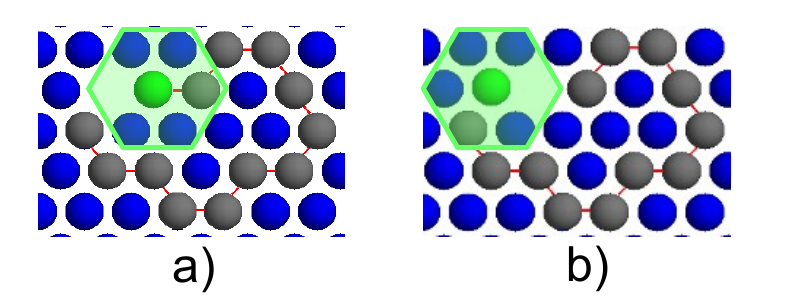} 	
\caption{ }
\label{metropolis}
\end{figure}

Fig.~\ref{metropolis}: Nearest neighbors using to calculate energy of sliding segment (green) before (a) and after (b) move of a single segment.

\begin{figure}[ht]
\includegraphics[scale=1.5]{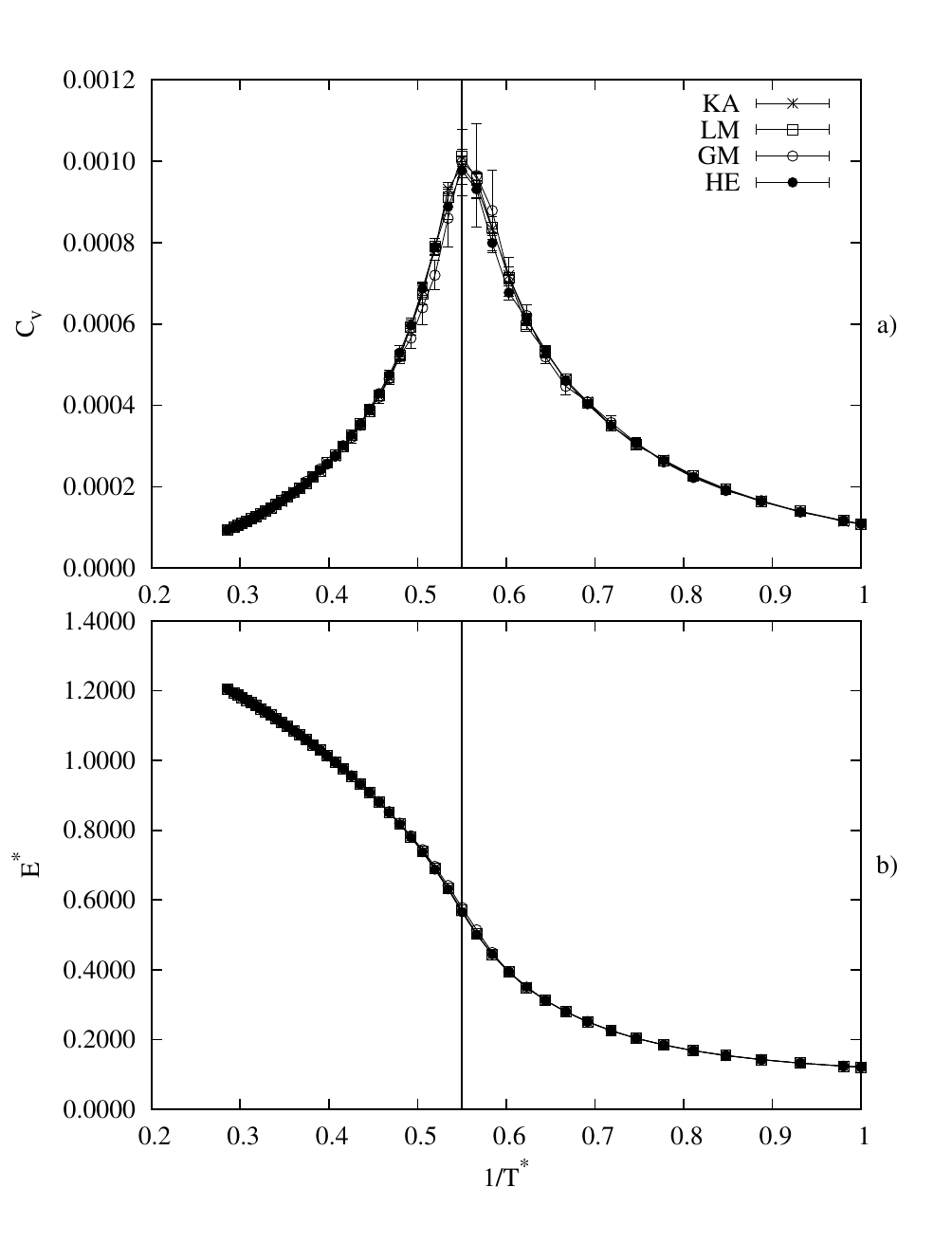} 	
\caption{ }
\label{fig1}
\end{figure}

Fig.~\ref{fig1}: Results of simulations (averaged over 5 runs) for $50\times 50$ lattice size and four different method: (a) heat capacity, $C_v$, with error bars (b) energy per lattice site, $E^*$. 

\begin{figure}[ht]
\includegraphics[scale=1.5]{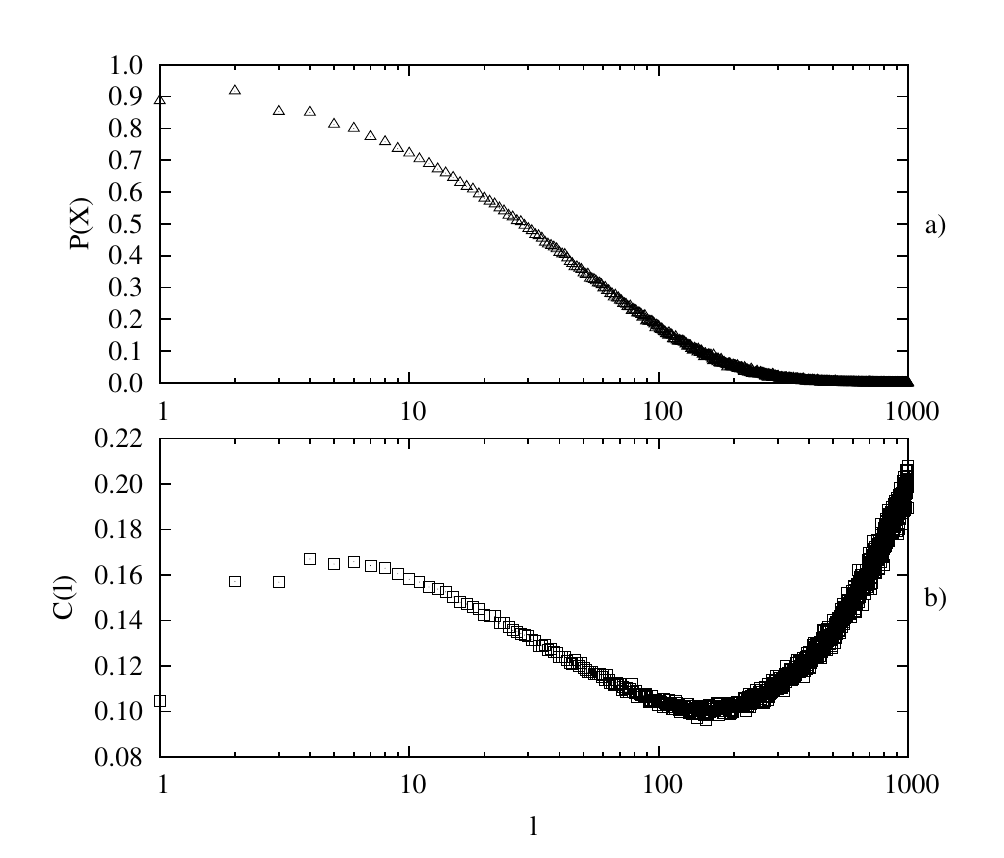} 	
\caption{ }
\label{fig2}
\end{figure}

Fig.~\ref{fig2}: Probability of loop acceptance, $P(X)$, (a) and contribution of different length of loop into simulation, $C(l)$, (b) in term of loop length, $l$, for the $GM$ method at $T^*=1.82$.

\begin{figure}[ht]
\includegraphics[scale=2.0]{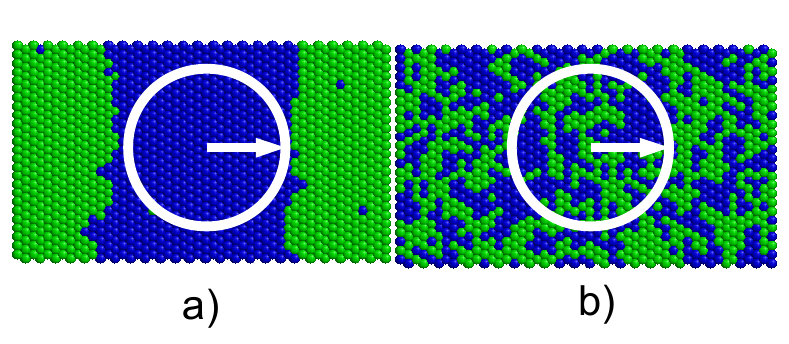} 	
\caption{ }
\label{fig3}
\end{figure}

Fig.~\ref{fig3}: Size of area (equal to $0.3L$) which is used in simulation to calculation the density profiles.

\begin{figure}[ht]
\includegraphics[scale=2.0]{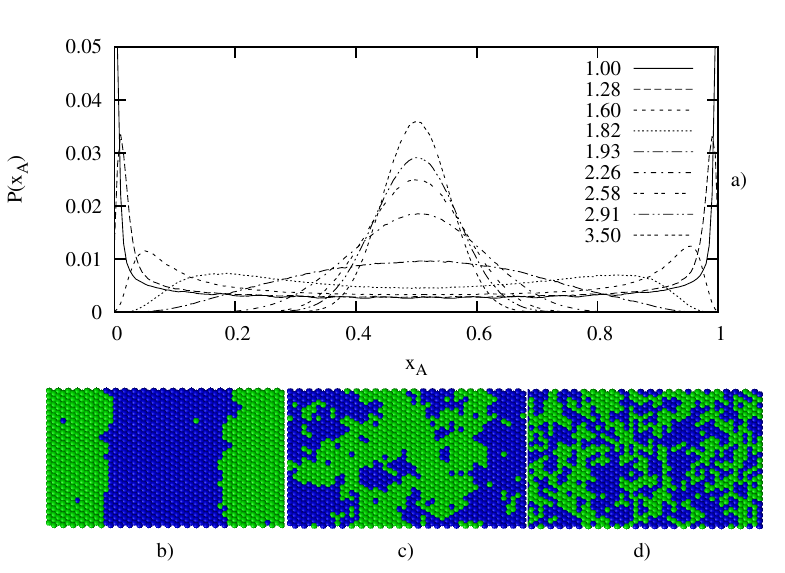} 	
\caption{ }
\label{fig4}
\end{figure}

Fig.~\ref{fig4}: Density profiles for $9$ different temperatures (a) and snapshots from simulation for $50\times 50$ lattice size at $T^* = 1.0$ (b), $1.82$ (c), $3.5$ (d).

\begin{figure}[ht]
\includegraphics[scale=2.0]{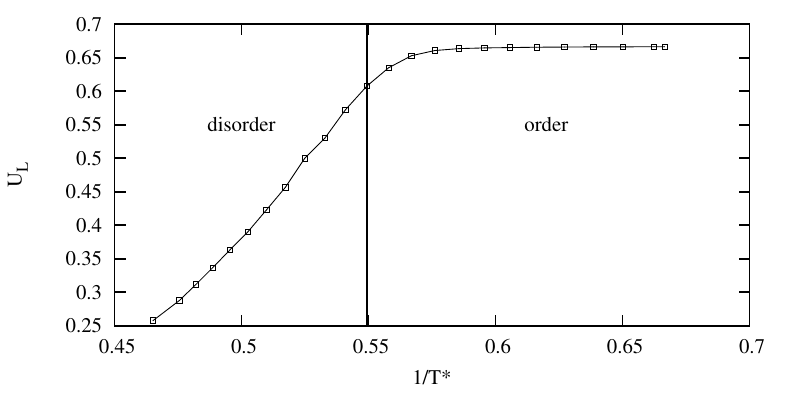} 	
\caption{ }
\label{fig4b}
\end{figure}

Fig.~\ref{fig4b}: The sample Binder cumulant in terms of $1/T^*$ for lattice size $100\times 100$.

\begin{figure}[ht]
\includegraphics[scale=1.5]{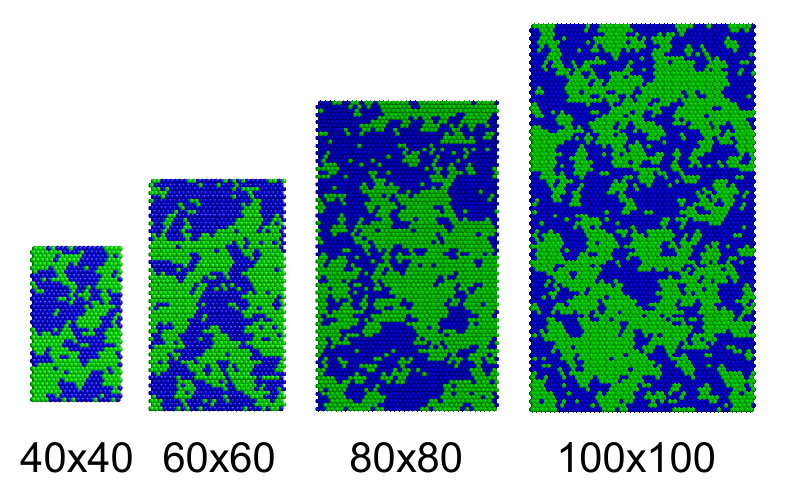} 	
\caption{ }
\label{fig4c}
\end{figure}

Fig.~\ref{fig4c}: Four sizes of box used in simulations to calculate the Binder cumulant.

\begin{figure}[ht]
\includegraphics[scale=2.0]{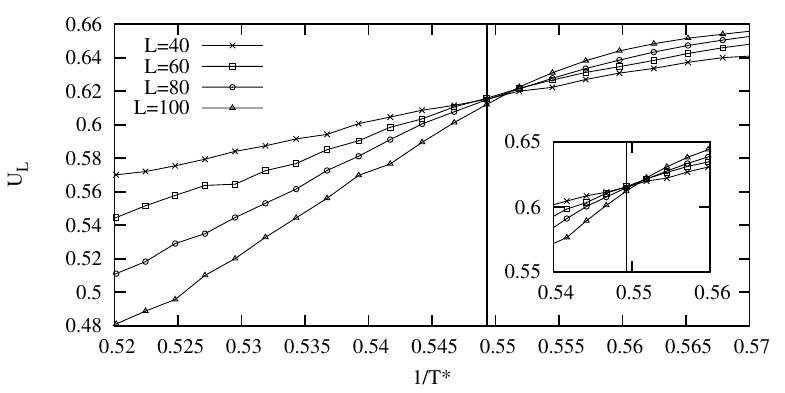} 	
\caption{ }
\label{fig5}
\end{figure}

Fig.~\ref{fig5}: The Binder cumulant (average over 3 runs) for the Kawasaki method and four different size of box. The inset shows intersection of the Binder cumulant.

\begin{figure}[ht]
\includegraphics[scale=2.0]{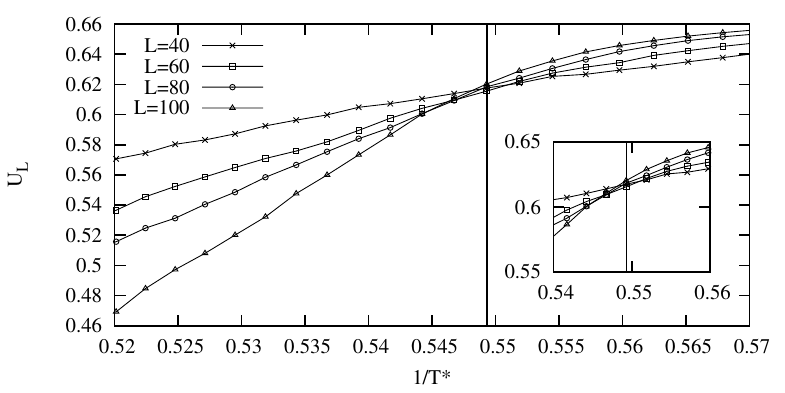} 	
\caption{ }
\label{fig6}
\end{figure}

Fig.~\ref{fig6}: The Binder cumulant (average over 3 runs) for the $LM$ method and four different size of box. The inset shows intersection of the Binder cumulant.

\begin{figure}[ht]
\includegraphics[scale=2.0]{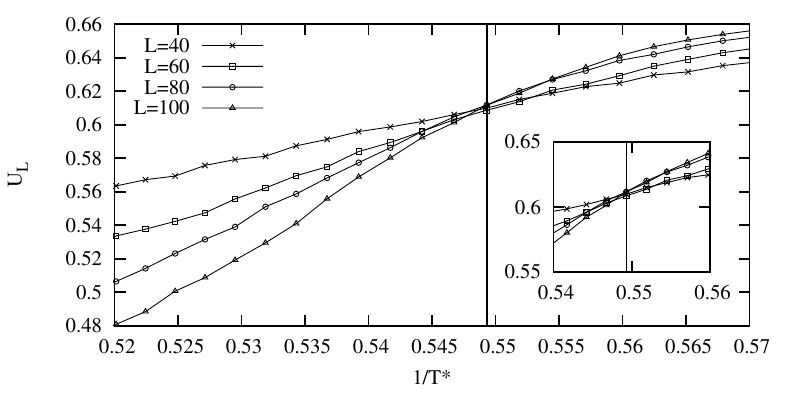} 	
\caption{ }
\label{fig8}
\end{figure}

Fig.~\ref{fig8}: The Binder cumulant (average over 3 runs) for the heuristic method and four different size of box. The inset shows intersection of the Binder cumulant.

\clearpage

\clearpage

\clearpage

\clearpage

\clearpage

\clearpage

\clearpage

\clearpage

\clearpage

\clearpage

\clearpage

\end{document}